\documentclass[prl,twocolumn,superscriptaddress,floatfix,letterpaper]{revtex4}
\usepackage{graphicx,amssymb,bm,natbib,amsfonts,amsmath,graphics,color}

\newcommand {\bisco}{Bi$_2$Sr$_2$CaCu$_2$O$_{8+\delta}$}
\newcommand {\LSCO}{La$_{2-x}$Sr$_x$Cu$_2$O$_{4}$}

\newcommand {\SLbisco}{Bi$_2$Sr$_{1.6}$La$_{0.4}$Cu$_2$O$_{6+\delta}$}

\newcommand {\degree}{$^{\circ}$}

\begin{document}
\title{Bond stretching phonon softening and angle-resolved photoemission kinks in optimally doped Bi$_2$Sr$_{1.6}$La$_{0.4}$Cu$_2$O$_{6+\delta}$ superconductors}

\author{J. Graf} \affiliation{Materials Sciences Division, Lawrence Berkeley National Laboratory, Berkeley, CA 94720, USA}
\author {M. d'Astuto}
\affiliation{Institut de Min\'eralogie et de Physique des
Milieux Condens\'es (IMPMC), Universit\'e Pierre et Marie Curie -
Paris 6, case 115, 4, place Jussieu, 75252 Paris cedex 05, France}
\altaffiliation{Institut de Min\'eralogie et de Physique des Milieux
Condens\'es (IMPMC), CNRS UMR 7590, Campus Boucicaut, 140 rue de
Lourmel, 75015 Paris, France}
\author{C. Jozwiak} \affiliation{Department of Physics, University of California Berkeley, CA 94720, USA} 
\author {D.R. Garcia}
\affiliation{Materials Sciences Division, Lawrence Berkeley National Laboratory, Berkeley, CA 94720, USA}
\affiliation{Department of Physics, University of California Berkeley, CA 94720, USA}\date {\today}
\author {N. L. Saini}
\affiliation {Dipartimento di Fisica, Universit\`a di Roma La Sapienza, 00185 Roma, Italy}
\author {M. Krisch}
\affiliation {European Synchrotron Radiation Facility, F-38043 Grenoble C\'edex, France}
\author{K. Ikeuchi}
\affiliation {SPring-8/JAEA,1-1-1 Kouto, Sayo, Hyogo, 670 Japan}
\author {A.Q.R. Baron}
\affiliation {SPring-8/RIKEN\&JASRI, 1-1-1 Kouto, Sayo, Hyogo, 670 Japan}
\author {H. Eisaki}
\affiliation {AIST Tsukuba Central 2, Umezono, Tsukuba, Ibaraki, 305-8568, Japan}
\author{A. Lanzara}\email{ALanzara@lbl.gov} \affiliation{Materials Sciences Division, Lawrence Berkeley National Laboratory, Berkeley, CA 94720, USA} \affiliation{Department of Physics, University of California Berkeley, CA 94720, USA}\date {\today}
\date {\today}
\begin {abstract}
We report the first measurement of the optical phonon dispersion in optimally doped single layer Bi$_2$Sr$_{1.6}$La$_{0.4}$Cu$_2$O$_{6+\delta}$ using inelastic x-ray scattering. We found a strong softening of the Cu-O bond stretching phonon at about $\mathbf{q}=(\approx 0.25,0,0)$ from 76 to 60 meV, similar to the one reported in other cuprates.  A direct comparison with angle-resolved photoemission spectroscopy measurements taken on the same sample, revealed an excellent agreement in terms of energy and momentum between the ARPES nodal kink and the $\textit{soft}$ part of the bond stretching phonon.  Indeed, we find that the momentum space where a 63$\pm$5 meV kink is observed can be connected with a vector $\mathbf{q}=(\xi,0,0)$  with $\xi\geq 0.22$, which corresponds exactly to the soft part of the bond stretching phonon mode. This result supports an interpretation of the ARPES kink in terms of electron-phonon coupling.
\end {abstract}
\maketitle

\paragraph*{}
\label{sec:INTRODUCTION}

The coupling of electrons to phonons can be studied from the renormalization of the phonon dispersions using both inelastic neutron (INS) and x-ray scattering (IXS). For example, if the phonon wavevector matches 2$\bm{k}_F$, where $\bm{k}_F$ is the Fermi momentum, an enhancement of the electron-phonon coupling occurs due to Fermi surface nesting, resulting in phenomena such as Kohn anomaly or charge density wave. 
Similarly, the coupling of electrons to thermal excitations as phonons or magnons is manifested in ARPES data by a kink in the electronic dispersion near the excitation frequency \cite{Ashcroft76}.  Hence a full mapping of both electron and phonon dispersion is fundamental to uncover the electron-phonon interaction in a material. 

INS and IXS data have shown an anomalous softening of the Cu-O bond stretching (BS) mode at the metal-insulator phase transition in several cuprates superconductors \cite{Pintschovius91,Uchiyama04,Pyka93,dAstuto02,McQueeney99,Reznik06,Chung03,Pintschovius99} as well as  non-superconducting perovskites \cite{Pintschovius89,Reichardt99,Reznik06}. Whether this anomalous softening is related to superconductivity or to the strong charge order instabilities present in some of these compounds is not clear yet, since neither of these two phenomena are ubiquitous to all of these compounds.

ARPES data have shown a kink in the electronic dispersion at 60-70 meV along the nodal direction in various p-type cuprates \cite{Lanzara01}. This kink shifts toward lower binding energy (30-40 meV) as the Brillouin zone face is approached \cite{Cuk04,Gromko03,Terashima07}.  
Due to the coinciding energy, the nodal kink has been associated with the BS phonon \cite{Lanzara01}. However, an alternative explanation has been suggested \cite{Johnson01,Kaminski01}.  Therefore, what is the true nature of the nodal kink, what determines the momentum region where it exists and what are the signatures of the anomalous phonon softening on the electronic structure are still open questions.

A combination of IXS and ARPES measurements on the same material is therefore very valuable in establishing, if any, the exact relation between these two anomalous behaviors.  Such a study has been missing so far. The best system to perform this cross analysis is the single layer \SLbisco\ superconductor (Bi2201) as no report of a magnetic resonance mode exists, allowing one to draw a direct comparison between the electronic structure and the phonon spectra.
Also, the Bi-based cuprates are among the most studied materials by ARPES because of the good quality of their cleaved surface.  On the other hand, the lack of large single crystals has made INS experiment difficult and, though IXS can probe sub-millimeter crystals, this is a challenging experiment due to the very low inelastic cross section of the BS mode \cite{dAstuto02b}.

We report the first measurements of the dispersion of the longitudinal phonons in optimally doped Bi2201.  We find a strong anomaly and renormalization of the BS phonon dispersion along the $\bm{q}=(\xi,0,0)$ direction, similar to the one reported in several cuprates \cite{Pintschovius91,Uchiyama04,Reznik06,Pyka93,dAstuto02,Chung03}. ARPES measurements revealed the presence of two distinct kink structures near the node and the BZ face at 60-70 and 30-40 meV respectively, similar to the two kink structure reported in double layer \bisco\ \cite{Cuk04,Gromko03} and \LSCO\ \cite{Terashima07}.

From the comparison between ARPES and IXS measurements performed on the same sample, we show that the $\emph{soft}$ part of the BS phonon matches the energy and momentum of the 60-70 meV kink.

\paragraph*{}
\label{sec:EXPERIMENTAL TECHNIQUE}
The single crystals of Bi2201 (T$_c$=33 K) were grown using the traveling-solvent floating-zone technique \cite{Eisaki04}.
The IXS experiment was performed at beamline BL35XU at SPring-8 in Japan \cite{Baron00} and completed at beamline ID28 at the European Synchrotron Radiation Facility in France \cite{Krisch02}. 
The momentum and energy resolutions were better than 0.08 \AA$^{-1}$ and 3.2 meV respectively. The measurements were done in the transmission geometry at low temperature ($\approx$10 K) to reduce contribution from the low energy phonons peaks. The sample thickness was $\approx$ 30$\mu$m, allowing $\approx$ 30\% of x-ray transmission.

IXS energy scans were carried out at $\bm{Q=G+q}$ points of the reciprocal lattice, where $\bm{G}$ is the zone center vector, and $\bm{q}$ is the reduced vector which corresponds to the phonon propagation vector. The exact $\bm{Q}$ points are:  $\bm{Q}$=(3.02,0,0.18),(3.09,0,0.06),(3.15,0,-0.06),(3.22,0,-0.19) at SPring8 and $\bm{Q}$=(3.25,0,0),(3.45,0,0) at ESRF. For the rest of the paper, we will neglect the c-axis component of the momentum transfer since the phonon anomaly was found to be independent of the c-component \cite{Reznik07}. The six $\bm{Q}$ points will be described by the reduced vector $\bm{q}=(\xi,0,0)$. 

The ARPES experiments were carried out at beamline V-4 of the Stanford synchrotron radiation laboratory with 22.4 eV photon. The ARPES data along the node were reproduced using an in-house 6 eV laser-based setup similar to the one described in ref. \onlinecite{Koralek06}. In both case the total energy and angular resolution were better than 14 meV and 0.4\degree, respectively and the sample temperature was bellow 15 K.

\paragraph*{}
\label{sec:RESULTS}
\begin{figure} \includegraphics[width=8cm]{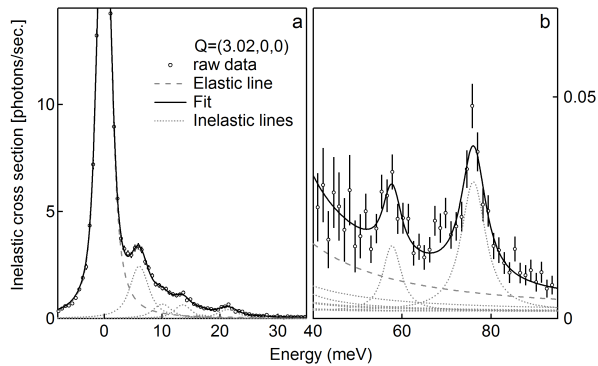}
\caption{IXS phonon spectrum of Bi2201 at 10K. The markers are the raw data, the dashed line is the resolution function and the plain line is a fit. (a) Low energy part of the spectrum. (b) High energy part with a smaller y-scale.}\end{figure}

Fig. 1 shows a typical IXS energy-scan in longitudinal geometry at $\bm{Q}$=(3.02,0,0).  A harmonic oscillator convoluted with the instrumental resolution function was used for the fit. Three features can be clearly distinguished in panel (a): (I) the elastic peak at 0 meV is the static diffuse scattering coming from disorder or weakly ordered structures; (II) the first longitudinal acoustic phonon centered around 6 meV and (III) two low energy optical modes at 14 and 21 meV. Other phonon modes are not resolved due to the dominating contribution from the tails of the elastic and acoustic signals. Panel (b) shows the same energy scan at higher energy. The small error bars above 50 meV reflect the very good statistical quality of the data. Two peaks corresponding to the last two longitudinal optical (LO) phonon modes are clearly resolved in the data.


\begin{figure}\includegraphics[width=8.5cm]{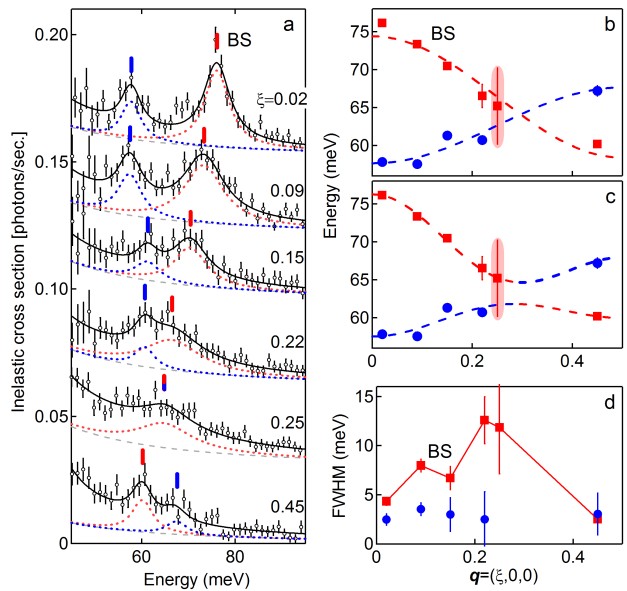}
\caption{(Color online) LO phonons dispersion. (a) IXS spectra for $\bm{Q}$=(3+$\xi$,0,0) with $\xi$ spanning from the BZ center (top spectrum, $\xi=0.02$) to the BZ face (bottom spectrum, $\xi=0.45$). The spectra are vertically shifted. The plain lines show the harmonic oscillator fit, the dashed lines show the elastic tail and the dotted lines show the last two mode used in the fit. (b,c) Phonon dispersions. The cosine dashed lines are guides for the eyes illustrating the crossing (b) and anti-crossing (c) scenarios. (d) FWHM. The error bars are an estimate of the standard deviation of the fit coefficients}\end{figure}

Fig. 2 shows the evolution of these two highest LO phonons across the Brillouin zone (BZ).  Panel (a) shows the IXS spectra for $\xi$ spanning from the BZ center to the BZ face. At the BZ center, two peaks are resolved. The first peak is at 58 meV and the second is at 76 meV. Both are almost resolution limited. As we approach the zone face, both peaks disperse toward 65 meV and as $\xi$ reaches 0.22-0.25, the two modes are not distinguishable despite the very good energy resolution used here.  The $\xi$=0.25 spectrum shows actually a rather ill-defined broad peak, extending far in the low energy side. For $\xi$=0.45 we can resolve again two sharp peaks.

In panel (b) and (c), we show the dispersion of the two branches across the BZ.  There are two likely scenarios depending on the symmetry of the two branches. If they have the same symmetry, they anti-cross, otherwise they simply cross.  The dashed lines are guides for the eyes used to illustrate the crossing (b) and the anti-crossing (c) scenarios. In the case of an anti-crossing, the character of the two modes are exchanged between the two branches at the anti-crossing point. Therefore in either case, the 60 meV peak at $\xi$=0.45 has the character of the Cu-O BS mode, leading to a total softening of 16 meV.  A similar anti-crossing phenomenon is observed in the n-doped cuprate Nd$_{1.86}$Ce$_{0.14}$CuO$_{4+\delta}$ \cite{dAstuto02,Braden05} where the Cu-O BS mode anti-crosses with the Nd-O BS mode. In the case of optimally doped \LSCO\ (LSCO), a mode with little dispersion is seen at the same energy, 58 meV and has been attributed to the O bond bending mode \cite{McQueeney99}.

In order to distinguish between the two scenarios, we undertook conventional phonon calculations using a classical shell model \cite{Mirone}.
The parameters used are based on the model from ref. [\onlinecite{Kovaleva04}]. This model did not reproduce the low and high energy modes observed experimentally in a reliable way. This indicates that spectral ellipsometry data alone could not provide enough constraints for the parameters of the model. All our attempts to find a better set of potentials compatible with previous established potential sets \cite{Chaplot95} failed.
Several factors might be responsible for such disagreement, including: (I) the charge distribution and bonds of the high Z atoms containing f electrons (Bi) are very difficult to model accurately \cite{Renker96}, and (II) the one-dimensional incommensurate superstructure along the b-axis \cite{Kirk88} which is neglected in the calculation.

Panel (d) shows the corresponding FWHM of each peak. The two modes are sharp and almost resolution limited across the whole BZ except for $\xi$=0.22-0.25, where only one broad peak is resolved.

\begin{figure} \includegraphics[width=8.5cm]{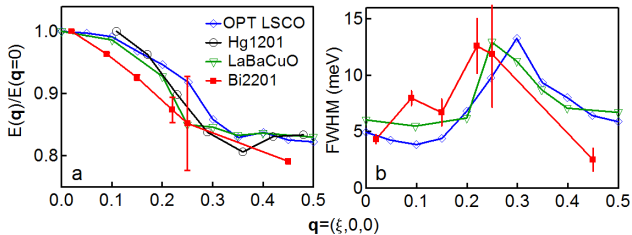}
\caption{\label{fig:3}(Color online) (a) Dispersion of the BS mode for various optimally p-doped single layer cuprates in comparison with the Bi2201 data. The frequencies are normalized to their value at $\xi=0$. The error bars are an estimate of the standard deviation of the fit coefficients.  The optimally doped LSCO and LaBaCuO are from \cite{Reznik07} and the Hg1201 are from \cite{Uchiyama04}. (b) FWHM when available.} \end{figure}

Fig. 3 shows a comparison of the Cu-O BS phonon dispersion (a) and FWHM
(b) for various optimally p-doped single layer cuprates. For Bi2201, an additional data point at $\xi\approx$0.3 is still needed to establish experimentally the exact momentum value of the maximal softening, though these data already suggest that the maximum FWHM is observed around $\xi\approx$0.22-0.25 in agreement with other cuprates \cite{Braden05}. We note however that the softening in the Bi2201 dispersion is smoother and observable already at very low $\xi$ in contrast with other cuprates.


\begin{figure} \includegraphics[width=8.5cm]{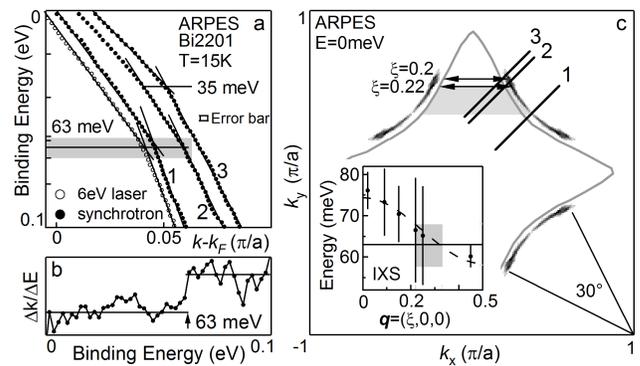}
\caption{\label{fig:4} (a) Electron energy vs momentum dispersion relation measured by ARPES for three different momentum cuts. Cut 1 is along the node and was measured with 22.4 eV synchrotron radiation (plain circles) and with 6 eV radiation using a laser based setup (open circles). Cut 2,3 are further toward the BZ face, close to the edge of the Fermi arcs reported in pseudogap state \cite{Kondo07}. Cut 2,3 were measured with synchrotron light only. The spectra are shifted horizontally and a $\approx$4 meV gap for cut 2,3 is subtracted. (b) First derivative of cut 1. (c) Experimental Fermi surface with the momentum location of the 3 cuts. The apparent finite Fermi arcs are due to the experimental resolution. The plain line shows a constant energy contour at the kink energy, 63meV. The inset of panel (c) shows the IXS dispersion and peak FWHM (error bars) of the BS mode discussed above. The shadow area represents the nodal kink energy and the $\bm{Q}$ vectors connecting the Fermi surface segments where the nodal kink is observed. Note that $\xi$=0.2 corresponds to 0.4 $\Delta k_x$ $(\pi/a)$.}\end{figure}

In Fig. 4, we compare the IXS and ARPES results for the phonon and electron dispersions in Bi2201. Panel (a) shows the energy momentum dispersion relation near the Fermi energy measured by ARPES along the nodal direction from the node to the BZ face. The exact momentum location of the cuts is shown in panel (c).  
In agreement with previous reports, we observe a kink at 63$\pm$5 meV along the nodal direction (cut 1). For cut 2, we observe in addition to the kink at 63meV, a kink at 35meV. For cut 3, we observe only the kink at 35 meV. The kink positions is extracted using a two straight lines fit \cite{Lanzara01}, the first derivative of the dispersion (panel b) and a self-energy type of analysis \cite{Verga03}. A similar shift of the kink toward lower binding energy has been reported for other cuprates  \cite{Cuk04,Gromko03,Terashima07}. The momentum where the 63 meV kink disappears seems to coincide with the tip of the Fermi arcs determined from the high temperature data \cite{Kondo07}. In this paper we will refer to the length of the Fermi arcs as determined from data in the pseudogap phase \cite{Kondo07}.

The direct comparison with the dispersion of the BS phonon clearly shows that the energy of the 63 meV kink coincides with the energy of the soft part of the BS phonon. To extend the comparison also to the momentum space we show in panel (c) a constant energy map of ARPES spectral intensity vs momentum at the Fermi energy (Fermi surface) in the superconducting state and a constant energy contour (gray line) at the nodal kink energy (63meV).

The comparison between the momentum of the BS phonon and the constant energy maps can provide unique information on the wavevectors needed for the electrons at the nodal kink to be scattered to the Fermi level by this particular phonon. The gray shadow area represents the range of $\bm{q}$=($\xi$,0,0) vectors that connect the electronic excitations at the 63meV kink energy (grey closed contour) to the Fermi surface segments from the node to the BZ face at the locations where the nodal kink is observed. For clarity, the gray area is shown only for half of a Fermi arc, but the same construction can be applied to the entire Fermi arc by symmetry.

We find that the entire momentum region where we observe a 63 meV kink can be connected with a momentum transfer $\bm{q}$=($\xi$,0,0) or equivalently  $\bm{q}$=(0,$\xi$,0) with $\xi\geq 0.22$ to an opposite, quasi-parallel Fermi surface segment. This momentum transfer corresponds exactly to the momentum region where the Cu-O BS phonon shows a softening and an anomalously {\it broad} lineshape (illustrated with the large error bars in the inset of panel (c)). Note that the furthest tips of the Fermi surface arcs are connected by $\bm{q}$=(0.5,0,0) which still corresponds to the soft BS mode. However, because of the relative orientation of these Fermi arcs (the momentum transfer is almost parallel to the Fermi arcs), the coupling with a perpendicular phonon ($\bm{q}$=(0,0.22,0)) is expected to be stronger. Interestingly, the maximum broadening corresponds to the case in which the momentum transfer is almost perpendicular to the Fermi arcs, suggesting that the anomalous softening and broadening observed in the phonon spectra might be a consequence of this Fermi surface topology.  
Note that although the weak doping dependence of the phonon softening seems to be in contrast with the strong doping dependence of the Fermi arcs, the broadening of the phonon peak at $\bm{q}$=0.25 is strongly doping dependent in line with the proposed scenario. Whether this scenario holds for the 35 meV mode as well, and the details of the doping dependence shall be address in further IXS and ARPES study.

Finally we note that the BS mode is supposed to be non-dispersive at about 80 meV along the other directions and in particular along [110] for the full breathing mode \cite{Pintschovius91,Braden05,Reznik07}. The absence of any strong feature above 63 meV in the ARPES data shows that the nodal charge carriers are preferentially coupled with the soft Cu-O half-breathing BS mode dispersing along the [100] direction as suggested by recent LSDA+U results \cite{Zhang07}. 
 
\paragraph*{}
\label{sec:SUMMARYANDCONCLUSIONS}
In conclusion, we report the first evidence of an anomalous dispersion of the Cu-O bond stretching phonon mode in a Bi-cuprate. A direct comparison with ARPES data shows that the energy and momentum where the strongest coupling to a 63 meV mode is observed corresponds exactly to the energy and momentum of the soft and broad Cu-O BS phonon.  Considering that a magnetic resonance mode was never observed in single layer Bi2201 \cite{sato03}, this result provides a strong support for the lattice origin of the 60-70 meV ARPES nodal kink. Although the role of electron-phonon coupling for superconductivity is not known yet, it is beyond any doubt that it is an important interaction linked with the Fermi surface topology, and it should not be neglected in any realistic theory of cuprates.

\begin{acknowledgments}
We thank D. Reznik and D.-H. Lee for useful discussions. The ARPES measurements were supported by the Division of Materials Sciences and Engineering, Office of Basic Energy Sciences of the U.S. Department of Energy under Contract No. DE-AC03-76SF00098. The IXS measurements were equally supported by the France-Berkeley grant, the Division of Materials Sciences and Engineering, Office of Basic Energy Sciences of the  U.S. Department of Energy under Contract No. DE-AC03-76SF00098 and by the by the National Science Foundation through Grant No. DMR-0349361.
\end{acknowledgments}

\end{document}